\documentclass[a4paper, amsfonts, amssymb, amsmath, reprint, showkeys, nofootinbib, twoside]{revtex4-1}
\usepackage[english]{babel}
\usepackage[utf8]{inputenc}
\usepackage[colorinlistoftodos, color=green!40, prependcaption]{todonotes}

\usepackage{amsthm}
\usepackage{mathtools}
\usepackage{physics}
\usepackage{xcolor}
\usepackage{graphicx}
\usepackage[left=23mm,right=13mm,top=35mm,columnsep=15pt]{geometry} 
\usepackage{adjustbox}
\usepackage{placeins}
\usepackage[T1]{fontenc}
\usepackage{lipsum}
\usepackage{csquotes}

\usepackage{algorithm}
\usepackage{algpseudocode}
\algnewcommand\algorithmicswitch{\textbf{switch}}
\algnewcommand\algorithmiccase{\textbf{case}}
\algnewcommand\algorithmicassert{\texttt{assert}}
\algnewcommand\Assert[1]{\State \algorithmicassert(#1)}%
\algdef{SE}[SWITCH]{Switch}{EndSwitch}[1]{\algorithmicswitch\ #1\ \algorithmicdo}{\algorithmicend\ \algorithmicswitch}%
\algdef{SE}[CASE]{Case}{EndCase}[1]{\algorithmiccase\ #1}{\algorithmicend\ \algorithmiccase}%
\algtext*{EndSwitch}%
\algtext*{EndCase}%

\usepackage[pdftex, pdftitle={Article}, pdfauthor={Author}]{hyperref} 
\begin{document}
\title{Efficient parallel algorithms for free-energy calculation of millions of water molecules in the fluid phases.}

\author{Luis Enrique Coronas}
    \affiliation{Departament de F\'isica de la Mat\`eria Condensada, Facutat de F\'isica, University of Barcelona, Spain}
    \affiliation{Institute of Nanoscience and Nanotechnology (IN2UB), University of Barcelona, Spain }

\author{Oriol Vilanova}
    \affiliation{Departament de F\'isica de la Mat\`eria Condensada, Facutat de F\'isica, University of Barcelona, Spain}
    \affiliation{Institute of Nanoscience and Nanotechnology (IN2UB), University of Barcelona, Spain }

\author{Giancarlo Franzese}
    \email[Correspondence email address: ]{gfranzese@ub.edu}
    \affiliation{Departament de F\'isica de la Mat\`eria Condensada, Facutat de F\'isica, University of Barcelona, Spain}
    \affiliation{Institute of Nanoscience and Nanotechnology (IN2UB), University of Barcelona, Spain }


\begin{abstract}
Simulating water droplets made up of millions of molecules and on timescales as needed in biological and technological applications is challenging due to the difficulty of balancing accuracy with computational capabilities. Most detailed descriptions, such as {\it ab initio}, polarizable, or rigid models, are typically constrained to a few hundred (for {\it ab initio}) or thousands of molecules (for rigid models). Recent machine learning approaches allow for the simulation of up to 4 million molecules with {\it ab initio} accuracy but only for tens of nanoseconds, even if parallelized across hundreds of GPUs. 
In contrast, coarse-grained models permit simulations on a larger scale but at the expense of accuracy or transferability. Here, we consider the CVF molecular model of fluid water, which bridges the gap between accuracy and efficiency for free-energy and thermodynamic quantities due to i) a detailed calculation of the hydrogen bond contributions at the molecular level, including cooperative effects, and ii) coarse-graining of the translational and rotational degrees of freedom of the molecules. The CVF model can reproduce the experimental equation of state and fluctuations of fluid water across a temperature range of 60 degrees around ambient temperature and from 0 to 50 MPa. In this work, we describe efficient parallel Monte Carlo algorithms executed on GPUs using CUDA, tailored explicitly for the CVF model. We benchmark accessible sizes of 17 million molecules with the Metropolis and 2 million with the Swendsen-Wang Monte Carlo algorithm.
\end{abstract}

\keywords{Fluid water, Thermodynamics, Metropolis Monte Carlo, Swendsen-Wang Monte Carlo, GPU paralellization}

\maketitle

\section{Introduction}
\label{section:Introduction}

Large-scale water modeling plays an essential role in simulations of biological systems and technological applications, where the balance between the model's accuracy and computational efficiency is crucial. On the one hand, a faithful representation of water properties is necessary to successfully reproduce the thermodynamic behavior of the entire system \citep{Chaplin}. On the other hand, the computational cost of modeling thousands or millions of water molecules, including explicit water-solute interactions, limits the accessible length and time scales of the simulation \citep{Onufriev:2018vj}. 

A detailed approach to simulate water systems is {\it ab initio} molecular dynamics (AIMD), which treats the nuclei classically while treating the electrons quantum mechanically. For a long time, this technique has been limited to systems of up to a few hundred molecules. However, thanks to recent advances in machine-learned DeepMD models, it is now feasible to simulate homogeneous nucleation with {\it ab initio} accuracy in systems of around hundreds of thousands of water molecules doi:\cite{PiaggiPNAS2022}. To our knowledge, the most extensive system benchmarked with this method was composed of 4 million water molecules, requiring parallelization over 480 - 27360 GPUs in the Summit supercomputer \citep{LU2021107624}. Although the length scale makes this approach promising for studying biochemical reactions, its computational cost limits the simulations to a few tens of ns, a short timescale for many biochemical and nanotechnological applications.

The accessible timescales can be extended by using models that represent water molecules at a lower level of description. The atomistic rigid TIP4P/2005 and the polarizable AMOEBA best describe the behavior of the systems 
\citep{KlesseJACS2020}, but they are typically limited to thousands of molecules and hundreds of nanoseconds. Alternatively, coarse-graining (CG) strategies reduce computational costs by averaging over the degrees of freedom that are believed to have a minor impact on the system's behavior. Among the most popular CG models used in biological simulations are MARTINI \citep{Tsanai2021} and SIRAH \citep{Machado2019, Klein2021}. MARTINI maps four water molecules into a single bead that interacts through effective potentials (4:1). Instead, SIRAH employs a mapping ratio of (11:4). However, at this level of description, these models cannot accurately reproduce hydrogen-bond (HB) interactions or cooperative effects \citep{Barnes1979}. Therefore, they leave the relevance of these interactions in biological systems unaddressed.  

The quest for a water model that simultaneously offers a detailed description of the HB network, including cooperativity, while also being suitable for large-scale simulations remains unresolved. In this context, the model proposed by Franzese and Stanley (FS) for water monolayers \citep{FSJPCM2002, FS2007, KFS2008, Mazza2011, Stokely2010, FranzeseSCKMCS2008, Santos:2011aa, delosSantos2012, BiancoSR2014, Bianco:2019aa, CoronasBook2022} stands out as a promising approach. 

The FS model describes the monolayer HB network at a molecular resolution, incorporating many-body contributions \citep{Stokely2010} while coarsening the translational degrees of freedom of the molecules through a discrete density field. It is suitable for long-time and large-scale simulations \citep{Mazza2011}, even under supercooled conditions \citep{BiancoSR2014, Bianco:2019aa}. Furthermore, its extension by Bianco and Franzese (BF), which includes the effect of interfaces, has been applied to biological problems such as protein folding \citep{BiancoPRL2015, Bianco:2017ab}, protein design \citep{BiancoPRX2017}, and protein aggregation \citep{BiancoJPhysChemLett2019, Bianco:2020aa, polym13010156}. In these studies, the BF model has helped to reveal the role of HB interactions in the complex behavior of proteins under various thermodynamic conditions.

Coronas, Vilanova, and Franzese (CVF) recently extended the FS model to bulk \citep{CoronasThesis, Coronas-2024, coronas2024phase}. They also demonstrated its applicability to hydrated biological interfaces \citep{CoronasThesis}.

Specifically, in Ref. \citep{Coronas-2024}, we showed that--thanks to a parametrization based on quantum {\it ab initio} calculations and experimental data--the model is thermodynamically {\it reliable}. It reproduces the experimental equation of state of water and thermodynamic fluctuations with outstanding accuracy. The range of quantitative agreement extends over 60 degrees, around 300 K at ambient pressure, and up to 50 MPa. 

In Ref. \citep{coronas2024phase}, we demonstrated that the CVF model is {\it transferable} to deep supercooled conditions, where it exhibits a liquid-liquid critical point in the thermodynamic limit. This finding is consistent with results obtained from optimized rigid models, such as rigid TIP4/Ice, and polarizable models, such as iAMOEBA.

In this work, we design parallel algorithms to show that the CVF model is also {\it scalable} and {\it efficient} for conducting large-scale simulations. To illustrate its scalability, we present results from simulations involving up to 17 million water molecules in the liquid phase. For these simulations, we needed only a few hours on a single workstation to calculate the thermodynamic properties at specific temperatures and pressures. To achieve this efficiency, we developed {\it in-house} software, which we describe here and offer as {\it open access} for further use and modifications by the scientific community.

Our code uses CUDA, a C-style programming language for kernels executed by the graphics processing unit (GPU) \citep{NVIDIA}. Over the last decade, CUDA has been widely utilized in Computational Physics to simulate, for example, lattice spin models using local and cluster Monte Carlo (MC) \citep{Hawick2011, Weigel2011, KomuraOkabe2012a}, molecular engines \citep{Hall2014}, Brownian motors \citep{Spiechowicz2015}, and to solve stochastic differential equations \citep{Januszewski2010}. GPU architectures are particularly effective for enhancing the performance of MC dynamics for spin models on regular lattice \citep{Hawick2011}. As we will demonstrate in the following sections, this is also true for the CVF model, which employs an underlying lattice structure to coarse-grain the density field and define the HB network.

We define both local and cluster MC algorithms for the CVF model. In both cases, we use the specific topological properties of our model. Consequently, both algorithms are tailored to the CVF model. However, our work may inspire the development of parallel algorithms for other models, such as those proposed in Refs. \cite{Borick:1995aa, Guisoni:2006oq, 10.1063/1.2434974, Fiore:2009aa, Bertolazzo:2014aa, PhysRevE.98.032116, cerdeirina2019} for water, or the model for ion hydration proposed by Dutta et al. \citep{Dutta2015}.

The model presented here is limited to the liquid phases of water, including supercooled states, and does not address the crystalline phases. This limitation arises from coarse-graining the coordinates of the molecules using a density field defined at the lattice resolution, where each cell's volume corresponds to the proper volume of the molecules. Consequently, we cannot define structural functions, such as the radial distribution function or the structure factor, to distinguish between the ice and fluid phases of water. This limitation will be addressed in the future by extending the model to incorporate the coordinates of the molecules, as done in Ref. \cite{Bianco:2014fk}, where polymorphism and melting via a hexatic phase were studied for a monolayer.

Additionally, for numerical efficiency, we do not permit molecular diffusion. Therefore, we cannot compute translationally dynamic quantities, such as the diffusion coefficient or characteristic translational decorrelation times. However, these quantities can be easily calculated within the framework of MC simulation by considering diffusive MC dynamics, as illustrated in Refs. \cite{FdlSJPCM2009, Santos:2011aa, delosSantos2012}, which allows us to estimate the occurrence of glassy dynamics and diffusive anomalies.

Nevertheless, our work allows unprecedented large-scale calculations for the thermodynamic observables of water in the fluid phases, including the supercooled region, while maintaining a detailed description of the HB network with quantitative precision. We pave the way for realistic simulations of large protein systems in explicit solvent, incorporating the effects that stem from individual HBs and their cooperativity.

The paper is organized as follows: 
In Section \ref{section:MaterialsMethods}, we present the model and define the algorithms for the local (Metropolis) MC and the cluster (Swendsen-Wang) MC calculations. 
In Section \ref{section:Results}, we show and discuss the results regarding critical slowing down in the supercooled liquid region, explain how the cluster MC allows us to overcome this issue, and provide a benchmark for the algorithm. 
In Section \ref{section:conclusion}, we address the advantages and limitations of our approach and present our conclusions. 
Technical details about the algorithms, the benchmarks and the workstation are provided in the appendices.

\section{Materials and Methods}
\label{section:MaterialsMethods}

\subsection{The model}
\label{subsection:Model}

We consider $N$ water molecules at a constant temperature $T$ and pressure $P$ within a fluctuating volume $V(T, P)$. We decompose the total volume $V$ into two components: homogeneous and heterogeneous. The homogeneous component, $V_{\rm iso}(T, P)$, corresponds to the isotropic contribution arising from the van der Waals interaction between the water molecules. This component is modeled using the Lennard-Jones potential with a hard-core distance $r_0$ and a cutoff at $6r_0$, where $r_0 \equiv 2.9$~\AA~ is the van der Waals diameter of a single water molecule, associated to its van der Waals volume $v_0$, as determined from experiments \citep{Finney:2001dq}. The cutoff is chosen large enough to include all significant contributions to the van der Waals interactions. Simultaneously, the truncation of the Lennard-Jones potential at the hard-core distance is introduced to simplify the model's implementation. Previous calculations~\citep{Santos:2011aa} have demonstrated that this truncation does not qualitatively impact the phase diagram of the monolayer (FS) model related to the current CVF model.
We take $\epsilon$, the characteristic energy of the Lennard-Jones interaction, as the internal unit of energy. From \emph{ab initio} energy calculations \citep{Henry2002}, we set $\epsilon \equiv 5.5$~kJ/mol.

The heterogeneous component of the volume reflects local fluctuations resulting from the formation of HBs under specific thermodynamic conditions. Sastry et al. demonstrated \citep{Sastry:1996aa} that assuming these fluctuations are proportional to the total number of HBs, $N_{\rm HB}$, is sufficient to reproduce water's volumetric anomalies. Thus, the total volume is expressed as
\begin{equation} 
\label{V} 
V \equiv V_{\rm iso} + v_{\rm HB} N_{\rm HB}, 
\end{equation} 
where the proportionality factor $v_{\rm HB}$ remains independent of the thermodynamic conditions $(T, P)$. This assumption is made to simplify the model and is shown to be reasonable {\it a posteriori}, at least within a limited range of $T$ and $P$. We will further discuss this limitation before the conclusions.

We set $v_{\rm HB} \equiv 0.6 v_0$. This choice stems from the volume difference per HB between high-density ice VI and VIII and low-density tetrahedral ice Ih \citep{BiancoSR2014, CoronasBook2022}. It is based on the reasonable assumption that the difference between the low- and high-density ices is solely due to the open structure associated with tetrahedral HB formation and that in low-density ice, all HBs are formed, with each water molecule engaging in four HBs. In contrast, in high-density ices, all HBs are absent.  

Our strategy for achieving large-scale simulation capability involves reducing the degrees of freedom of water without losing information about the HB network. To this end, we replace the atomic coordinates of the $N$ molecules with a density field defined at the resolution of a single molecule. Therefore, we partition the homogeneous component of the total volume $V_{\rm iso}$ into {\cal N} cells, each sized $v_{\rm iso} \equiv V_{\rm iso}/{\cal N} \geq v_0$, and ensure that ${\cal N} \geq N$. The case ${\cal N} = N$, examined here, corresponds to bulk water, with each cell accommodating a single molecule and $v_{\rm iso}$ representing the proper volume of a water molecule that forms no HBs.
The case ${\cal N} > N$ indicates that the volume is shared by water and solutes, which will be addressed in a future publication for the 3D case \citep{CoronasThesis}. In 2D, the case ${\cal N} > N$ has been considered already when there are vacancies in a monolayer \cite{FdlSJPCM2009, Santos:2011aa, delosSantos2012} or for hydrated proteins, e.g., in \cite{Fauli:2023aa} and references therein. 

To account for local volume fluctuations caused by the formation of  HBs, we associate a heterogeneous component $v_{\rm HB} n_{{\rm HB}, i}/2$ with each cell $i \in [1, \dots, {\cal N}]$. Here, $n_{{\rm HB}, i}$ represents the number of HBs formed by the molecule within cell $i$, with $\sum_i n_{{\rm HB}, i}/2=N_{\rm HB}$.
The factor of $1/2$ avoids double-counting of HBs. Consequently, each cell $i$ has a local volume defined as 
\begin{equation}
v_i\equiv v_{\rm iso}+v_{\rm HB} n_{{\rm HB},i}/2, 
\label{vi}
\end{equation}
that implies the Eq.(\ref{V}).

To keep our coarse-graining approach (i) straightforward to implement and (ii) consistent with both low and high coordination numbers in the fluid, we partition the total volume $V$ into cells of a cubic lattice. As a result, the relation between the van der Waals diameter $r_0$ and the associated volume is $v_0 \equiv r_0^3$. 

The partition of the volume $V$ into a cubic grid of cells is appropriate because DFT-based Car-Parrinello molecular dynamics simulations indicate that the water coordination number does not exceed six under ambient conditions \cite{Skarmoutsos:2022ur}. Simulations at high pressures also confirm this finding \cite{Saitta2003, Dietmar-Paschek:2008vl}. Therefore, each cell $i$ has six nearest neighbors.

\begin{table}[htp]
\caption{The CVF parameters. The parameters for the Lennard-Jones potential modeling the van der Waal interaction, $\epsilon$ and $r_0$, are adopted as units of energy and length, respectively.}
\begin{center}
\begin{tabular}{|c|c|c|c|c|c|c|}
\hline
$\epsilon$ & $r_0$ & cutoff & $v_0$ & $v_{\rm HB}$ & $J/(4\epsilon)$ &$J_\sigma/(4\epsilon)$ \\
\hline
$5.5$~kJ/mol& $2.9$~\AA & $6~r_0$ & $ r_0^3$ & $0.6 ~v_0$ &  $0.5$ & $0.08$\\
\hline
\end{tabular}
\end{center}
\label{parameters}
\end{table}%

The average distance between neighboring molecules is defined as $r \equiv v_{\rm iso}^{1/3} \in [r_0, \infty)$. Note that $r$ is unaffected by $N_{\rm HB}$, as the formation of HBs reduces the coordination number of water but 
does not alter the separation between molecules. Specifically, each water molecule minimizes the enthalpy of its local environment by forming four HBs in an almost perfect tetrahedral arrangement while excluding any 'interstitial' water molecule. This rearrangement leads to a decrease in local density, which corresponds with an increase in the effective volume of each molecule in the network, resulting in a variation in the total $V$ as described by Eq. (\ref{V}). 

The system is compressible, meaning that $V$ fluctuates at fixed $P$ in accordance with the equation of state. Therefore, for each $i$, the two components of $v_i$ ($v_{\rm iso}$ and $v_{\rm HB} n_{{\rm HB},i}/2$) vary. 
For the first, independent of $i$, it holds that $v_0/v_{\rm iso} = r_0^3/r^3 \in [0,1]$. 
By defining $r_{\rm 1/2}$ as the value where $r_0^3/r_{\rm 1/2}^3 = 0.5$, we classify as gas-like the cells with $r > r_{\rm 1/2}$, i.e., those with $v_{\rm iso} \geq 2 v_0$, and as liquid-like the others. 
Since this definition is independent of $i$, the entire system is either gas or liquid-like, depending on the value of  $V_{\rm iso}$. Nevertheless, local changes in the HB network, via $n_{{\rm HB},i}$, lead to heterogeneities in local volume fluctuations.

We consider negligible the HB formation in the gas and assume that molecules within gas-like cells cannot form HBs since the average O--O distance between them exceeds the HB-breaking threshold \citep{Luzar-Chandler96}. In addition, according to \emph{ab initio} simulations and the Debye-Waller factor \citep{Teixeira1990}, only $\widehat{OOH}$ angles (between two water molecules) within 60 degrees result in a bonded state. 
Therefore, only $1/6$ of the possible relative orientation states of two water molecules at HB distance can form a HB. To account for this, we introduce a bonding variable $\sigma_{ij} \in [1, \dots, q]$ with $q = 6$. It describes the relative orientation between molecules in neighboring cells $i$ and $j$. 
Each molecule $i$ has six bonding variables, $\sigma_{ij}$, one for each of the six neighboring molecules $j$.  A HB is formed when $\sigma_{ij} = \sigma_{ji}$, in such a way that a bonded state has a probability $1/q=1/6$ to occur. 

As discussed in \cite{coronas2024phase} and \cite{Coronas-2024}, the model splits the HB interaction into two components: (i) covalent (pairwise directional)  \citep{Shi:2018ac}, and (ii) cooperative (many-body) 
\citep{Barnes1979}, with a characteristic energies $J$ and $J_\sigma$, respectively. We set $J \equiv 11$~kJ/mol, i.e., $J/(4\epsilon) \equiv 0.5$, which is consistent with the energy of a single HB and cluster analysis \citep{Stokely2010}.

The cooperative HB interactions arise from many-body effects, contributing to the tetrahedral arrangement of HBs at low temperatures \citep{Cisneros2016}. The model includes interactions up to the five-body term within the first coordination shell, as derived from polarizable models \citep{Abella:2023ab}. Based on DFT calculations \citep{Cobar2012}, we set $J_\sigma/(4\epsilon) = 0.08$ to optimize the model’s accuracy in predicting the experimental equation of state and thermodynamic fluctuations \citep{Coronas-2024}.

The six bonding variables $\sigma_{ij}$ of the same molecule $i$ interact cooperatively, resulting in a reduction in energy $J_\sigma$ when $\sigma_{ij} = \sigma_{ik}$ for $j\neq k$. 
With a coordination number of six, this implies that the maximum cooperative energy in a cell is $15 J_\sigma$. For the selected model's parameters (Table \ref{parameters}), it follows that $J_\sigma \ll J$, which aligns with the general understanding that HB cooperative reorganization occurs at a temperature significantly lower than the formation of individual HBs \citep{Cisneros2016}. 

Both experimental and computational studies indicate that bulk water molecules form four tetrahedral HBs in their lowest energy state. Excited states correspond to defects in the HB network, where molecules have either fewer or more HBs. \emph{Ab initio} calculations for liquid water at ambient conditions show that under-coordinated molecules comprise approximately 45\% of the HB network, while over-coordinated molecules account for less than 5\% \citep{DiStasio:2014vu}. To simplify, we impose a maximum of four HBs per molecule, introducing variables $\eta_{ij}$, which are set to 1 or 0 depending on whether the HB between molecules $i$ and $j$ is allowed, as described in Appendix~\ref{CheckerboardPartition}.

Finally, the enthalpy of the system is given by:
\begin{equation} 
H(N, P, T) = U_{\rm LJ} - N_{\rm HB} J - N_{\sigma} J_{\sigma} + P V,
\label{H}
\end{equation} 
where $U_{\rm LJ}$ is the Lennard-Jones potential, $N_{\rm HB} \equiv \sum_{\langle i,j \rangle} \theta(r_{1/2} - r_{i,j}) \delta(\sigma_{i,j}, \sigma_{j,i}) \eta_{i,j}$, $N_{\sigma} \equiv \sum_k^N \sum_{\langle i,j \rangle} \delta(\sigma_{k,j}, \sigma_{k,i})$, and $V$ is given by Eq. (\ref{V}). Here, $r_{i,j} \equiv |\vec{r_i} - \vec{r_j}|$ is the distance between molecules $i$ and $j$, and $\theta(r)$ and $\delta(i,j)$ are the Heaviside step and Kronecker delta functions, respectively.
Thus, the formation of a macroscopic HB network leads to an increase in volume for the Eq. (\ref{V}), a decrease in entropy due to the reduced number of accessible $\sigma_{ij}$ states, and an increase in HB enthalpy, for the Eq. (\ref{H}).

\subsection{Monte Carlo step definition}
\label{MonteCarloStep}

A configuration of the CVF model is defined by the variables $\{\vec{r_i},\sigma_{ij},\eta_{ij}\}$. In the present version of the model, as discussed above, we coarse-grain the molecule position $\vec{r_i}$ over the lattice cell and assign to each molecule a proper volume $v_i$, Eq.(\ref{vi}).
Therefore, considering the definitions of $V_{\rm iso}$ and $\eta_{ij}$, the CVF configuration of the present model reduces to $\{V_{\rm iso},\sigma_{ij},\eta_{ij}\}$.

In each MC step, we update these variables in the following order:
\begin{enumerate}
 \item Update $V_{\rm iso}$, keeping $\{\sigma_{ij}, \eta_{ij}\}$ fixed. 
 \item Update $\eta_{ij}$, keeping $\{V_{\rm iso}, \sigma_{ij}\}$ fixed.
 \item Update $\sigma_{ij}$, keeping $\{V_{\rm iso}, \eta_{ij}\}$ fixed.
\end{enumerate}

We use the standard Metropolis algorithm to update the global variable $V_{\rm iso}$. This method involves accepting or rejecting a tentative change from $V_{\rm iso}$ to $V_{\rm iso} +\Delta V_{\rm iso}$ with a probability proportional to $\exp{[-P\Delta V_{\rm iso}/(k_B T)]}$, where $\pm \Delta V_{\rm iso}\propto \mathcal{O}(V_{\rm iso}/100)$.
We update the $\eta_{ij}$ as described in Appendix~\ref{CheckerboardPartition}. For the $\sigma_{ij}$, we employ the parallel local Metropolis algorithm (section \ref{subsection:Metropolis}) or the parallel Swendsen-Wang cluster algorithm (section \ref{subsection:Swendsen-Wang}), depending on the temperature: Metropolis for  $T \geq 208$~K 
 and  Swendsen-Wang for lower temperatures. 

\subsection{Metropolis}
\label{subsection:Metropolis}

\begin{figure*}
 \centering
 \includegraphics[width=0.8\textwidth]{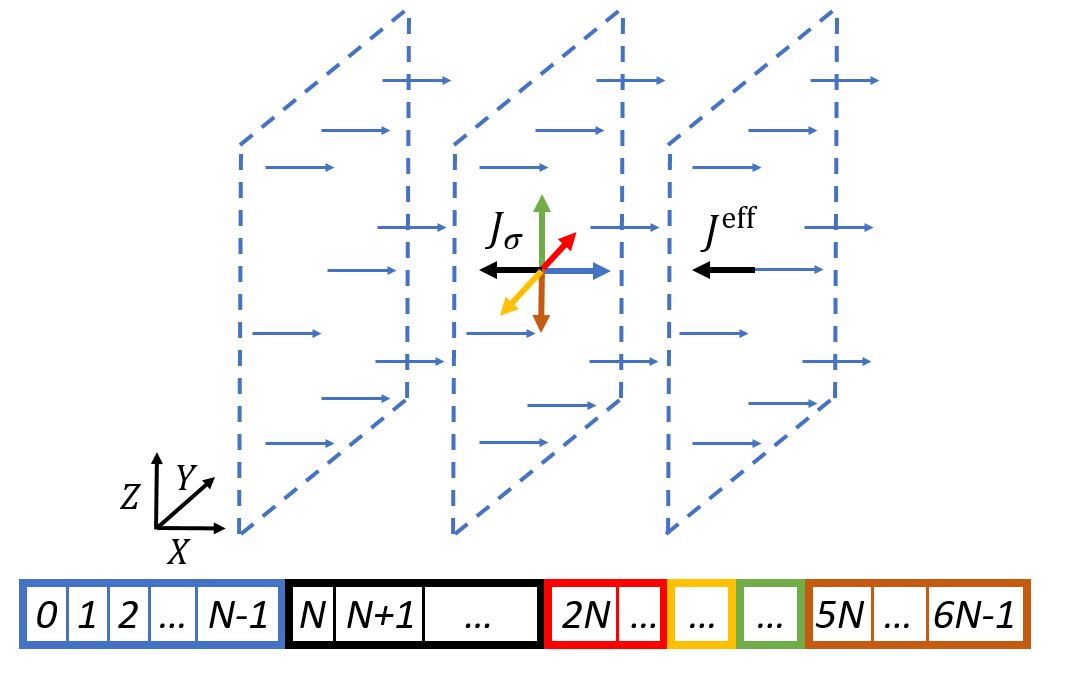}
 \caption{{\bf  Schematic illustration of the layered domains partitioning the bonding indices $\{\sigma_{ij}\}$.}
{\bf (Top)} Three layers (marked by dashed lines) of water molecules along the direction $x$. If $i$ is the central molecule, the colored arrows represent the bonding indices $\{\sigma_{ij}\}$ where $j$ runs over the n.n. molecules. 
The bonding index's color code with the n.n. molecule is blue for the right molecule, black for the left molecule, red for the back molecule, yellow for the front molecule, green for the top molecule, and brown for the bottom molecule. 
For clarity, we indicate only the blue arrow for the other molecules in the figure.
The set $\{\sigma_{ij}\}$ is divided into six domains, one for each color.
Therefore, the blue domain, represented in the figure, includes all the $\sigma_{ij}$ where $x_j=x_i+(v_{\rm iso}/v_0)^{1/d}$. 
The blue variable $\sigma_{ij}$ interacts with the five (different colors, same molecule) $\sigma_{ik}$ with $k\neq j$ via the cooperative interaction with characteristic energy $J_\sigma$, and, if $\eta_{ij}=1$,  with the black variable $\sigma_{ji}$ via the characteristic energy $J^{\rm eff} \equiv J- P v_{HB}$.
{\bf (Bottom)} Array sorting of the $\eta_{ij}$ and $\sigma_{ij}$ variables, according to the indexing formula described in the text, grouped by the color-coded domains. The ordering is relevant since it enables coalesced reading, improving the performance.}
 \label{fig:parallel_metropolis}
\end{figure*}

The Metropolis algorithm on a regular lattice can be efficiently parallelized by dividing the space into domains for simultaneous variable updates. To maintain detailed balance, the enthalpy change from altering a variable must remain independent of other variables within the same domain. For the Ising model, common partitioning schemes include layered \citep{Barkema1994} and checkerboard \citep{Heermann1990} methods, with CUDA implementations available for both 2D and 3D \citep{Hawick2011, Weigel2011, Wojtkiewicz2015}. However, these methods are not easily applicable to the CVF model due to differing lattice topologies, so we use a layered partition that enables memory coalescing in the CVF model.

We partition the $\{\sigma_{ij}\}$ variables, as described in Fig.~\ref{fig:parallel_metropolis} (Top), into six domains. Each domain contains variables interacting with six bonding indices: five on the same molecule and one on a n.n. molecule, all from different domains.
Therefore, we can update all the $\sigma_{ij}$ in the same domain simultaneously. 

In CUDA applications, the main bottleneck in execution arises from data access latency \citep{Tapia2011}. Performance can be enhanced by efficiently sorting memory to exploit memory coalescing \citep{Leist2009, Sanders2010, NVIDIA}. The GPU creates, manages, schedules, and executes blocks of 32 threads simultaneously, called \emph{warps} \citep{Hawick2011}. When a kernel reads (or writes) to global memory locations, it performs a single coalesced read (or write) transaction for every half-warp of 16 threads. Therefore, we are interested in sorting the vectors so that consecutive threads read (or write) consecutive memory addresses. 

We achieve this by sorting the arrays that store $\eta_{ij}$ and $\sigma_{ij}$ variables according to the index ${\tt idx} = {\tt arm} \cdot N + {\tt cell}$, where ${\tt arm} \in \{ 0, 1, ..., 5 \}$, and ${\tt cell} \in \{ 0, 1, ..., N-1 \}$. The index ${\tt arm}$ represents the six possible neighbors of the cell (from $0$ to $5$: left, right, front, back, top, bottom), and ${\tt cell} \equiv i$ is the index of the cell (Fig. ~\ref{fig:parallel_metropolis} Bottom).

We implement a CUDA kernel {\tt gpu\_metropolis(arm)} that launches one thread per water molecule $i$, where ${\tt arm}$ indicates which of the six independent domains is updated (Algorithm 1). We define a parallel Metropolis update as six sequential calls to {\tt gpu\_mertopolis(arm)}, where {\tt arm} is chosen randomly to mimic the random selection of $\sigma_{ij}$ variables in the sequential Metropolis and to avoid the propagation of correlation waves. 

We illustrate how the kernel {\tt gpu\_metropolis(arm)} performs coalesced memory transactions with the following example. We consider a half-warp that updates the block ${\tt dir}=0$ (left domain) of the water cells ${\tt cell}\in \{0, ..., 15\}$. Thus, {\tt idx} takes values from $0$ to $15$. When the kernel estimates $\Delta N_{\rm HB}$, it reads the \textit{right} arms of the neighboring cells $1$ to $16$, i.e., the (consecutive) positions {\tt idx} $=65$ to $80$. The same occurs when estimating $\Delta N_\sigma$, as the kernel reads memory positions in consecutive domains. We observe that an exception to this rule arises when the neighboring cell is positioned on the opposite side of the simulation box due to the periodic boundary conditions.

\subsection{Swendsen-Wang}
\label{subsection:Swendsen-Wang}

Local MC algorithms, such as Metropolis, experience a critical slowdown in their dynamics as the correlation length approaches the system size (as discussed in Section \ref{dynamics_metropolis}). In contrast, cluster MC algorithms efficiently update entire correlated regions of spins (clusters) simultaneously. Consequently, they produce statistically independent configurations at significantly lower computational costs. This efficiency is crucial in the supercooled region, for example, where the model exhibits a liquid-liquid phase transition culminating in a liquid-liquid critical point \citep{coronas2024phase}. In this context, we examine the Swendsen-Wang (SW) multi-cluster algorithm \citep{SwendsenWang}. 
The algorithm is defined so that, at each step, clusters of $\sigma_{ij}$ variables are formed with sizes ranging from 1 (an isolated $\sigma_{ij}$ variable) to the system's size $N$. This formation follows a distribution that reproduces that of thermodynamically correlated degrees of freedom, as discussed in detail in  \cite{Bianco:2019aa} based on site-bond correlated percolation \cite{Kasteleyn_Fortuin, PhysRevLett.42.518}.
The new configuration is generated by updating all the (correlated) $\sigma_{ij}$ variables within the same cluster to a new state. The sequential SW algorithm for the CVF model proceeds as follows:
\begin{enumerate}
 \item Visit all the cells $i$. For each $i$, loop over all the pairs of variables $(\sigma_{ij},\sigma_{ik})$. If they are in the same state, place a {\it fictitious} bond between them with probability $p_\sigma = 1-\exp\left( - J_\sigma/k_BT \right)$. 
 
 \item Visit all the pairs of n.n. cells $\langle i,j\rangle$. If $J^{\rm eff} \equiv J-P v_{HB} >0$, $\eta_{ij}=1$, and  $\delta_{\sigma_{ij},\sigma_{j,i}}=1$, place a {\it fictitious} bond with  probability $p_{\rm eff} = 1-\exp\left( - \left|J^{\rm eff}\right|/k_BT \right)$. Instead, if  $J^{\rm eff}<0$, place a {\it fictitious} bond with probability  $p_{\rm eff}$ if $\delta_{\sigma_{ij},\sigma_{ji}}=0$.  
 
 \item Use the Hoshen-Kopelman algorithm \citep{HoshenKopelman1976} to identify the clusters of $\sigma_{ij}$ variables connected by {\it fictitious} bonds.
 
 \item Visit all the clusters. For each, choose a random integer ${\tt rnd\_int} \in \{0, ..., q-1\}$. Change the state of all the $\sigma_{ij}$ variables in the cluster to  $\sigma_{ij} \leftarrow \left( \sigma_{ij}+ {\tt rnd\_int}\right)\%q$, where $\leftarrow$ is the assignment operator and $\%$ is the modulo operation.
\end{enumerate}

The SW algorithm performs three independent tasks. 
First, it places {\it fictitious} bonds between $\sigma_{ij}$ variables to generate the clusters. 
Second, it identifies all the clusters. 
Third, it updates each cluster. 
The first and third tasks are highly localized and can be easily parallelized. However, this is not the case for the cluster labeling operation. To tackle this challenge, we build on the work of Hawick et al., who developed various parallel labeling algorithms for arbitrary and lattice graphs using CUDA \citep{Hawick2010b}. Among these, the label equivalence algorithm was refined by Kalentev et al. \citep{Kalentev2011} and later applied by Komura and Okabe to SW simulations of the 2D Potts model \citep{KomuraOkabe2012a}. In this context, we modify the Hawick-Kalentev label-equivalence algorithm for the CVF model.

For a given SW step, we first generate the clusters. We directly parallelize this task so that each thread works on one CVF cell. Each thread is responsible for the cooperative interactions within its cell and the covalent interactions with the left, front, and top directions. To accomplish this, we allocate the array {\tt connected} of size $(15+3)N=18N$, which indicates whether two neighboring $\sigma_{ij}$ variables belong to the same cluster. We nest this array based on the index 
${\tt con\_idx} = {\tt link}N + {\tt cell}$, where ${\tt link} \in \{0, ..., 17\}$.
${\tt link}=0$, $1$, and $2$ represents the covalent connections between {\tt cell} and its neighbors in the left, front, and top directions. 
The values ${\tt link} \in \{3, ..., 17\}$ represent the 15 cooperative connections within ${	\tt cell}$.

Once the bonds are placed, we apply the label equivalence algorithm. We allocate the ${\tt label}$ array of size $6N$, which indicates the cluster that $\sigma_{ij}$ belongs to. Thanks to Kalentev's sophistication, this array also resolves label equivalences \citep{Kalentev2011}. The advantage is the reduction of the memory cost of the algorithm, which is significant due to the limited storage resources of the GPUs. We initialize ${\tt label}$ as ${\tt label}[{\tt idx}]={\tt idx}$, where ${\tt idx}$ is the $\sigma_{ij}$ index defined in \ref{subsection:Metropolis}. The algorithm resolves label equivalences through iterative calls to the \emph{scanning} and \emph{analysis} functions \citep{Kalentev2011, KomuraOkabe2012a}. When the algorithm converges, all the $\sigma_{ij}$ variables in the same cluster will take the same ${\tt label}$ value. 

The scanning function compares the label of a site ${\tt idx}$ to the labels of all the n.n. $\sigma_{ij}$ within the cluster. 
For every ${\tt idx}$, {\tt label}$[{\tt idx}]$ is updated to the minimum value among all the labels of the bonded sites, including itself. 
In Ref. \citep{KomuraOkabe2012a}, Komura and Okabe implemented this function using a single kernel for the 2D Potts model. 
However, for the CVF model, we find it more convenient to divide this function into two kernels. 
First, in {\tt gpu\_scanning\_covalent}, each thread scans left, front, and top covalent interactions (Algorithm 2).
Second, {\tt gpu\_scanning\_cooperative} scans the cooperative interactions (Algorithm 3).
An alternative implementation in a single kernel leads to race conditions when two threads attempt to update the same element of {\tt label}\footnote{To avoid this, we could use the CUDA {\tt atomic\_min} function; however, we found that it resulted in worse performance due to increased thread divergence \citep{Kalentev2011}.}.

Next, the analysis function updates {\tt label}$[{\tt idx}]$ (Algorithm 4). This step further propagates the minimum value of {\tt label} to other $\sigma_{ij}$ variables within the same cluster. Although the parallel implementation of the analysis function experiences race conditions, these collisions between threads will eventually be resolved in subsequent applications of the scanning and analysis functions \citep{Kalentev2011}. To minimize the impact of thread conflicts, we implement the {\tt gpu\_analysis}$({\tt arm})$ kernel, which updates only the {\tt label} of those $\sigma_{ij}$ variables in the ${\tt arm}$ domain. We then loop through the six domains to account for all the lattice sites.

To check whether the algorithm has converged, we first store a copy of the {\tt label} vector before calling the scanning and analysis functions, and then we compare it to the updated {\tt label}. We parallelize this task by assigning one thread to each CVF cell. The algorithm converges when the ${\tt label}$ remains unchanged. We provide an example of {\tt label} convergence after successive applications of the scanning and analysis functions in Table \ref{TableS1_SwendsenWang}.

\begin{table*}
\caption{{\bf Example of parallel label equivalence algorithm.} We consider a small cluster of seven $\sigma_{ij}$ variables with indices in ``$\sigma_{ij}$ index" row in a lattice of $N=64$ cells. Each pair of $\sigma_{ij}$ variables in the same cell (cell index~$\leftrightarrow $~Cartesian coordinates row) are bonded through a cooperative interaction. The pairs of $\sigma_{ij}$ variables $(0,65)$, $(257,337)$, and $(17,82)$ are bonded through a covalent interaction. The initial value of label coincides with the $\sigma_{ij}$ index. The following lines show the resulting label after the application of the kernels scan covalent, scan cooperative, and analysis. At the third iteration, {\tt label} does not change, so this cluster has converged. The SW step ends when all the clusters converge.}
\label{TableS1_SwendsenWang}
\centering
\begin{tabular}{ c|cc|cc|cc|c }
\hline
{\tt cell} index $\leftrightarrow (x,y,z)$& \multicolumn{2}{c|}{0 $\leftrightarrow $(0,0,0)} &\multicolumn{2}{c|}{1$\leftrightarrow $(1,0,0)} & \multicolumn{2}{c|}{17$\leftrightarrow $(1,0,1) } & 18$\leftrightarrow $(2,0,1)\\
$\sigma_{ij}$ index & 64 & 0 & 65 & 257 & 337 & 17 & 82 \\
\hline
initial label & 64 & 0 & 65 & 257 & 337 & 17 & 82 \\
\hline
scan covalent & 64 & 0 & 0 & 257 & 257 & 17 & 17 \\
scan cooperative & 0 & 0 & 0 & 0 & 17 & 17 & 17 \\
analysis & 0 & 0 & 0 & 0 & 17 & 17 & 17 \\
converged? & \multicolumn{7}{c}{No} \\
\hline
scan covalent & 0 & 0 & 0 & 0 & 0 & 17 & 17 \\
scan cooperative & 0 & 0 & 0 & 0 & 0 & 0 & 17 \\
analysis & 0 & 0 & 0 & 0 & 0 & 0 & 0 \\
converged? & \multicolumn{7}{c}{No} \\
\hline
scan $+$ analysis & 0 & 0 & 0 & 0 & 0 & 0 & 0 \\
 converged? & \multicolumn{7}{c}{Yes} \\
\hline
\end{tabular}
\end{table*}

\section{Results and discussion}
\label{section:Results}

\subsection{Critical slowdown of the Metropolis dynamics in the vicinity of the liquid-liquid critical point}
\label{dynamics_metropolis}

The CVF model predicts a liquid-liquid phase transition (LLPT) between high-density liquid (HDL) and low-density liquid (LDL) phases in the supercooled region \citep{coronas2024phase}. The LLPT ends in a liquid-liquid critical point (LLCP), located at $P_C =$ 174 $\pm$ 14 MPa and $T_C =$ 186 $\pm$ 4 K in the thermodynamic limit ($N\to\infty$). This is in close agreement with finite-$N$ estimates from iAMOEBA \citep{Pathak2016}, TIP4P/Ice \citep{Debenedetti289}, and ML-BOP \citep{Dhabal_Kumar_Molinero_2024} models, as well as with a recent estimate from a collection of experimental data \citep{Mallamace2024}. 

Approaching the LLCP, the correlation length $\xi$ of the water HB network increases and ultimately diverges at the critical point. Consequently, the autocorrelation time $\tau$ of the local Metropolis MC dynamics, which is proportional to $\xi$, also increases approaching the LLCP. This can be demonstrated by calculating the  autocorrelation function
\begin{equation}
 \label{eq:correlation}
 C_M(\Delta t) \equiv \frac{\langle M_i(t_0+\Delta t) M(t_0)\rangle-{\langle M \rangle}^2}{\langle M^2 \rangle - {\langle M \rangle}^2},
\end{equation}
where $M \equiv \frac{1}{qN} \max\{N_{q'}\}$ is an order parameter, and $N_{q'}$ is the number of $\sigma_{ij}$ variables in the state $q' \in \{0, ... , q-1 \}$. We define the autocorrelation time $\tau$ as the time at which $C(\tau) = 1/e$. Thus, $C(\Delta t)$ allows us to estimate the autocorrelation time $\tau$ of the HB network. Two CVF configurations are uncorrelated if they are sampled after a number of MC steps $\geq \tau$. 

\begin{figure*}
 \centering
 \includegraphics[scale=1]{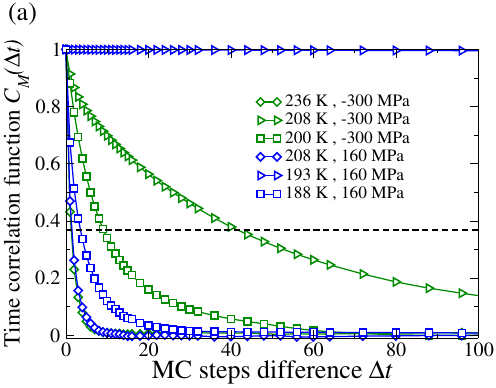}
 \includegraphics[scale=1]{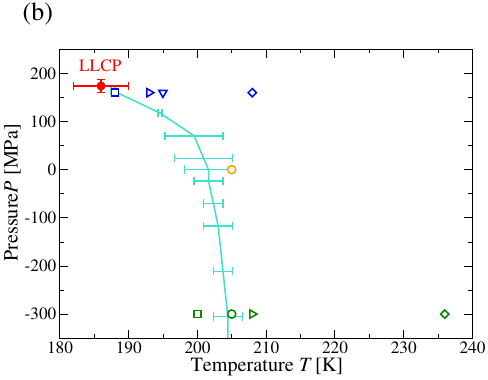}
 \caption{{\bf (a) Correlation function $\mathbf{C_M(\Delta t)}$ of the order parameter $\mathbf{M}$ for parallel Metropolis MC algorithm.} 
The time $\Delta t$ is measured in units of MC steps. 
The system size is $N=32\,768$. 
At $P=$160 MPa and $T=$193 K (blue triangles), near the estimate of the supercooled water LLCP in the thermodynamic limit 
\citep{coronas2024phase}, the correlation function decreases very slowly, consistent with the critical slowing down expected for local MC dynamics near a critical point. As temperature increases ($T=208$ K, blue diamonds) or decreases ($T=188$ K, blue squares) at constant pressure, the correlation decays more rapidly. A similar trend is observed at low-$P=-300$ MPa, with slow decay at $T=208$ K (green triangles) and faster decay at higher $T=236$ K (green diamonds) and lower $T=200$ K (green squares). The dashed line indicates the $C_M = e^{-1}$ value corresponding to the autocorrelation time $\tau$. {\bf (b) Location of simulated thermodynamic points in the $\mathbf{P}$-$\mathbf{T}$ phase diagram.} From the LLCP (red), the locus of extrema of the correlation length $\xi$, i.e., the Widom line, emerges. As a proxy estimate of the Widom line, we plot the locus of maxima of the specific heat, i.e., the maxima of enthalpy fluctuations (turquoise line) as discussed in \citep{coronas2024phase}. The blue and green symbols correspond to the thermodynamic conditions selected in panel (a) and Fig. \ref{fig:correlation_comparison}.}
 \label{fig:metropolis_correlation}
\end{figure*}

We calculate, with the Metropolis algorithm, $C_M(\Delta t)$ for a system with $N=32\,768$ water molecules (Fig.~\ref{fig:metropolis_correlation}.a).
At a pressure of $P = 160$ MPa, which is close to the critical pressure in the thermodynamic limit ($P=174\pm 14$ MPa), $\tau$ displays non-monotonic behavior with fast dynamics at high $T = 208$ K and low $T= 188$ K, alongside an apparent divergence at $T=193$ K.
This behavior is linked to a structural change between HDL-like and LDL-like forms of water, characterized by the Widom line (the locus of maxima of $\xi$) emerging from the LLCP at higher pressure \citep{coronas2024phase}. 
As an approximate estimate of the Widom line, we present the locus of extrema of the specific heat $C_P$, namely the maxima of the enthalpy fluctuations (Fig. \ref{fig:metropolis_correlation} b). At extremely low pressure, $ P = -300$ MPa, far from the critical region, $\tau$ exhibits a similar non-monotonic behavior, but without any apparent divergence upon crossing the Widom line. 
This is consistent with a decrease in the maxima of the correlation length $\xi$ as the distance between the LLCP and the point along the Widom line increases.

\subsection{Cluster MC dynamics avoids the critical slowdown}
\label{dynamics_cluster}

\begin{figure*}
 \centering
 \includegraphics[scale=1.1]{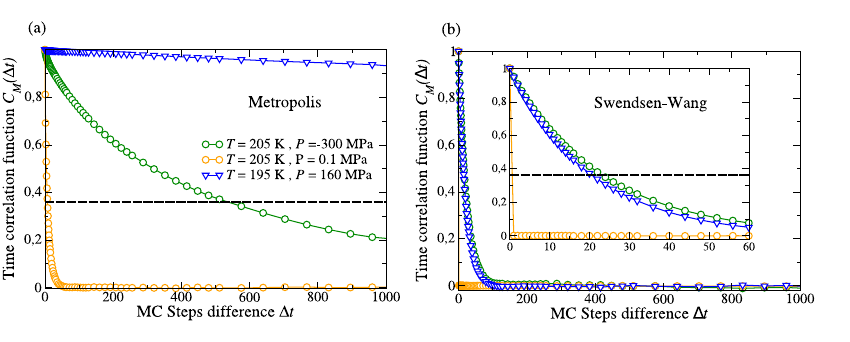}
 \caption{{\bf Comparison between correlation function $\mathbf{C_M(\Delta t)}$ computed with Metropolis and SW.} (a) For the Metropolis MC, at $P=$160 MPa and $T=$195 K (blue triangles), near the LLCP, the correlation function decreases very slowly, as discussed in Fig.~\ref{fig:metropolis_correlation}. 
At a higher temperature of $T=205$ K and a lower pressure of $P=-300$ MPa (green circles) near the Widom line, the correlation function exhibits a slow decay. 
Away from the Widom line, at the same temperature but at a higher pressure of $P=0.1$ MPa (orange circles), the correlation function decays much faster. 
(b) The SW algorithm avoids the critical slowdown of dynamics at the same state points (symbols and colors as in panel a). Inset: An enlarged view of the short-time regime enhances the distinction among the data for different state points.}
 \label{fig:correlation_comparison}
\end{figure*}

Cluster MC algorithms are suitable for efficiently sampling the critical region, as they bypass the critical slowdown of the dynamics by updating regions of correlated HBs simultaneously. We compare the autocorrelation function computed with local Metropolis (Fig. \ref{fig:correlation_comparison}.a) and cluster SW (Fig. \ref{fig:correlation_comparison}.b) algorithms. As discussed in Section \ref{dynamics_metropolis}, we observe slow dynamics of the system at low $P=-300$ MPa and $T=205$ K, near the Widom line. With increasing pressure ($P=0.1$ MPa) at constant temperature, the system remains in a metastable supercooled liquid state, exhibiting rapid decorrelation. Finally, at $T=195$ K and $P=160$ MPa, close to the LLCP (174 $\pm$ 14 MPa, 186 $\pm$ 4 K) \citep{coronas2024phase}, the autocorrelation time appears to diverge. The comparison with SW illustrates that cluster MC circumvents the critical slowdown of the dynamics in all cases, even near the LLCP.

\subsection{Benchmark of the algorithms}
\label{Methodology:Benchmarking}
As discussed above, the autocorrelation time $\tau$ for the SW algorithm is significantly shorter than that for the Metropolis MC. However, SW cluster MC is notably more computationally expensive than Metropolis. Therefore, to determine which MC dynamics is more efficient in generating uncorrelated configurations, one must compare the time each algorithm takes to produce $\tau$ MC steps.

First, we analyze the computational cost of the parallel Metropolis algorithm for different system sizes, $N\leq 17\, 576\, 000 $. The hardware and software specifications of the workstation are detailed in \ref{workstation}. Depending on the system size, we perform between $2$ and $10$ independent simulations of $5\,000$ MC steps. We find that the results are robust against changes in thermodynamic conditions; that is, changes in $T$ and $P$ do not affect the computational cost of the algorithm (Algorithm 1).

\begin{figure}
 \centering
 \includegraphics[width=0.48\textwidth]{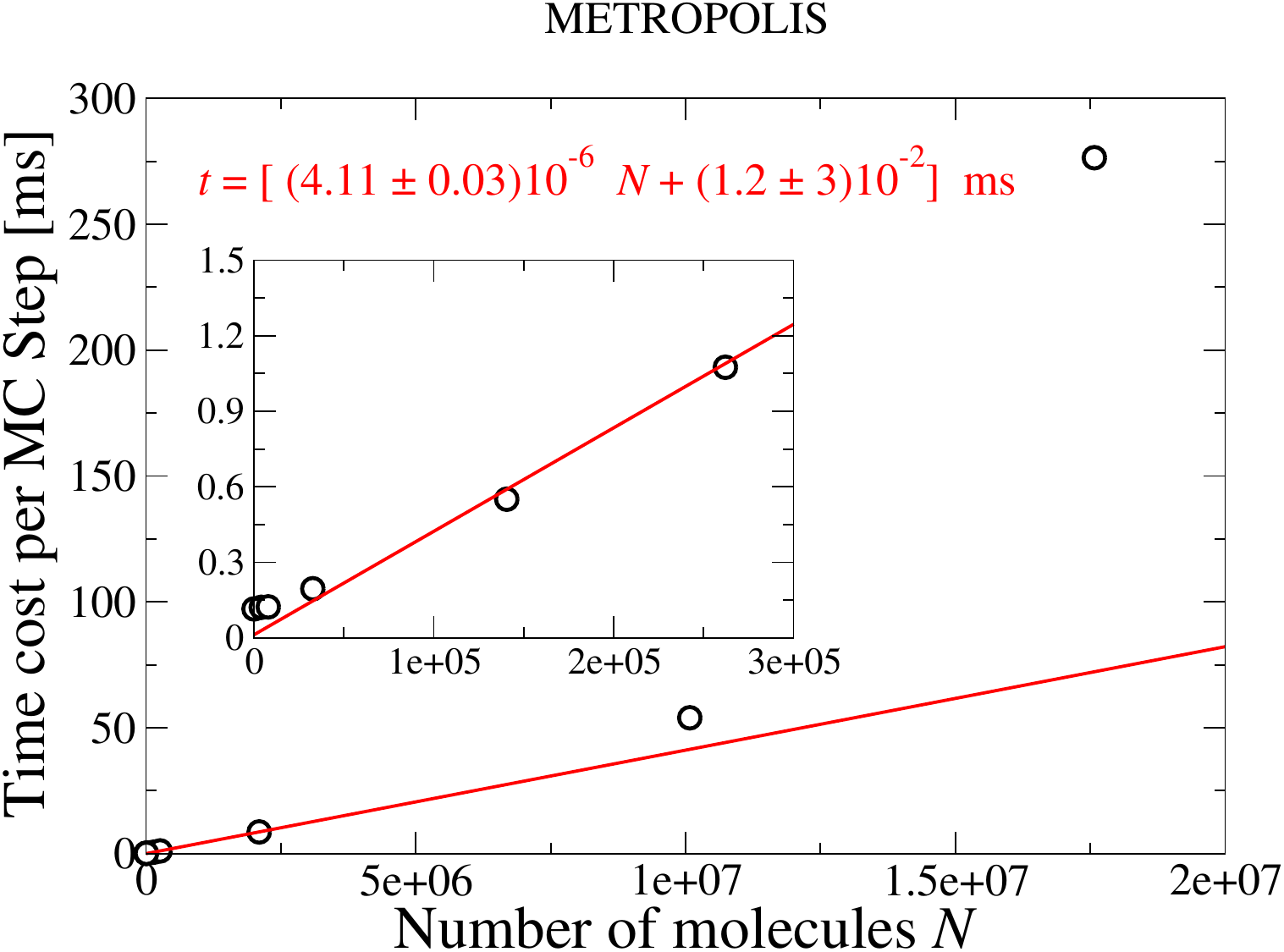}
 \caption{{\bf Time cost of a parallel Metropolis update of 64 $\leq \mathbf{N} \leq$ 17\, 576\, 000 water molecules}. The line is a linear fit $t=a N+b$ of the time cost within the range $32\,768 \leq N \leq 2\,097\,152$, with fitting parameters $a = (4.11 \pm 0.03)10^{-6}$~ms and $b= (1.2 \pm 3)10^{-2}$~ms. We observe a large deviation from linearity for $N>10^7$.
Inset: The enlarged view at small $N$ highlights the deviation from linearity for $N\leq 8\,000$, with a time cost saturation $\sim 0.1$~ms at small $N$.
In both the main panel and the inset, error bars are smaller than the size of the symbols.}
 \label{fig:MetroplisGPUtime}
\end{figure}

Our results show that the time necessary (cost) for a parallel Metropolis update scales linearly for $32\,768 \leq N \leq 2\,097\,152$ (Fig.~\ref{fig:MetroplisGPUtime}). For these systems, the GPU resources are neither saturated (large $N$) nor under-exploited (small $N$); thus, the time spent on data accessing scales linearly with $N$. For small $N\leq 8\,000$, the computational resources of the GPU are not optimized. We find that in this range, the time cost of a Metropolis step remains approximately constant ($\sim 0.1$~ms, Fig.~\ref{fig:MetroplisGPUtime}: inset).
For large $N\geq 2\,097\,152$, the size of the arrays of random numbers must be reduced to fit within the GPU global memory (Appendix~\ref{random_numbers}). The additional time cost arises from both the increasing number of executions of the kernels for generating random numbers and the time involved in memory transactions. In particular, we benchmark accessible size-systems up to $N=17\,576\,000$ water molecules with a time cost of $280$~ms per Metropolis update (Fig.~\ref{fig:MetroplisGPUtime}), which corresponds to a cubic simulation box of 75$\times$75$\times$75 nm$^3$.  

We estimate the size-dependent performance gain, or speedup factor, SF$(N)\equiv t^{\rm CPU}/t^{\rm GPU}$, defined as the ratio between the time required for a parallel and a sequential update of an entire system of $N$ molecules, as shown in Table \ref{table:MetropolisSpeedupFactor}. The results indicate that, for the smallest system considered ($N=64$ molecules), the parallel algorithm is less efficient than the sequential one. This is not surprising, as a sufficiently large number of threads must be executed to fully utilize the GPU resources~\citep{Hall2014}. Indeed, Wojtkiewicz and Kalinowski also find ${\rm SF} < 1$ for small systems~\citep{Wojtkiewicz2015}. The large SF$=136.72$ measured for $N=2\,097\,152$ is attributed to the significant increase in the time cost of the sequential implementation compared to the parallel approach. More specifically, we find that the large $\sigma$ and $\eta$ arrays exceed the RAM storage capacity, necessitating that they be loaded in portions, which delays the sequential computation. We could not measure the SF for $2\,097\,152 < N \leq 17\,576\,000$ due to the excessive cost of computing $t^{\rm CPU}$. Instead, we extrapolated SF$=208$ for $N=10\,077\,696$ and SF$=245.8$ for $N=17\,576\,000$ from a power-law fit in the range $32\,768 \leq N \leq 2\,097\,152$ (see Fig. \ref{fig:MetropolisSpeedupFactor}). Further details on the computation of the SF are provided in Appendix~\ref{speedup}.

\begin{table}
\centering
\caption{Speedup factor, SF$\equiv t^{\rm CPU}/t^{\rm GPU}$, of the GPU Metropolis algorithm in comparison to the sequential implementation on the CPU for $N$ water molecules. The error in the last digit of the estimate is indicated in parentheses. (*) For $N \geq 10\,077\,696$, we extrapolate SF from a power law fit (Fig \ref{fig:MetropolisSpeedupFactor}).}
\begin{tabular}{ |c|c|}
\hline
\multicolumn{2}{|c|}{Metropolis Speedup Factor (SF)}\\
\hline
&\\[-1em]
 Number of molecules $N$ & SF\\
\hline
 64 & 0.1159(6) \\
\hline
 4\,096 & 7.09(3) \\
\hline
8\,000 & 13.42(9) \\
\hline
 32\,768 & 37.8(3) \\
\hline 
 140\,608 & 63.8(3) \\
\hline
 262\,144 & 63.5(3) \\
\hline
 2\,097\,152 & 136.72(3) \\
\hline
 10\, 077\, 696 (*) & 208.0 \\
\hline
 17\, 576\, 000 (*) & 245.8 \\
\hline
\end{tabular}
\label{table:MetropolisSpeedupFactor}
\end{table}

Next, we estimate the performance of the parallel SW algorithm. Unlike the Metropolis case, the time cost of a SW update depends on the cluster size distribution, which in turn is influenced by the thermodynamic conditions \citep{Bianco:2019aa}. Close to the Widom line, the system undergoes a transition from non-percolation to percolation upon isobaric cooling. Thus, we consider two temperatures on either side of the transition: $T=$ 195~K (percolation) and 210~K (not percolation), with $P=0.1$~MPa. For every system size $N$, we perform between 5 and 10 independent simulations of $10^3$ MC steps.

\begin{figure}
 \centering
 \includegraphics[width=0.48\textwidth]{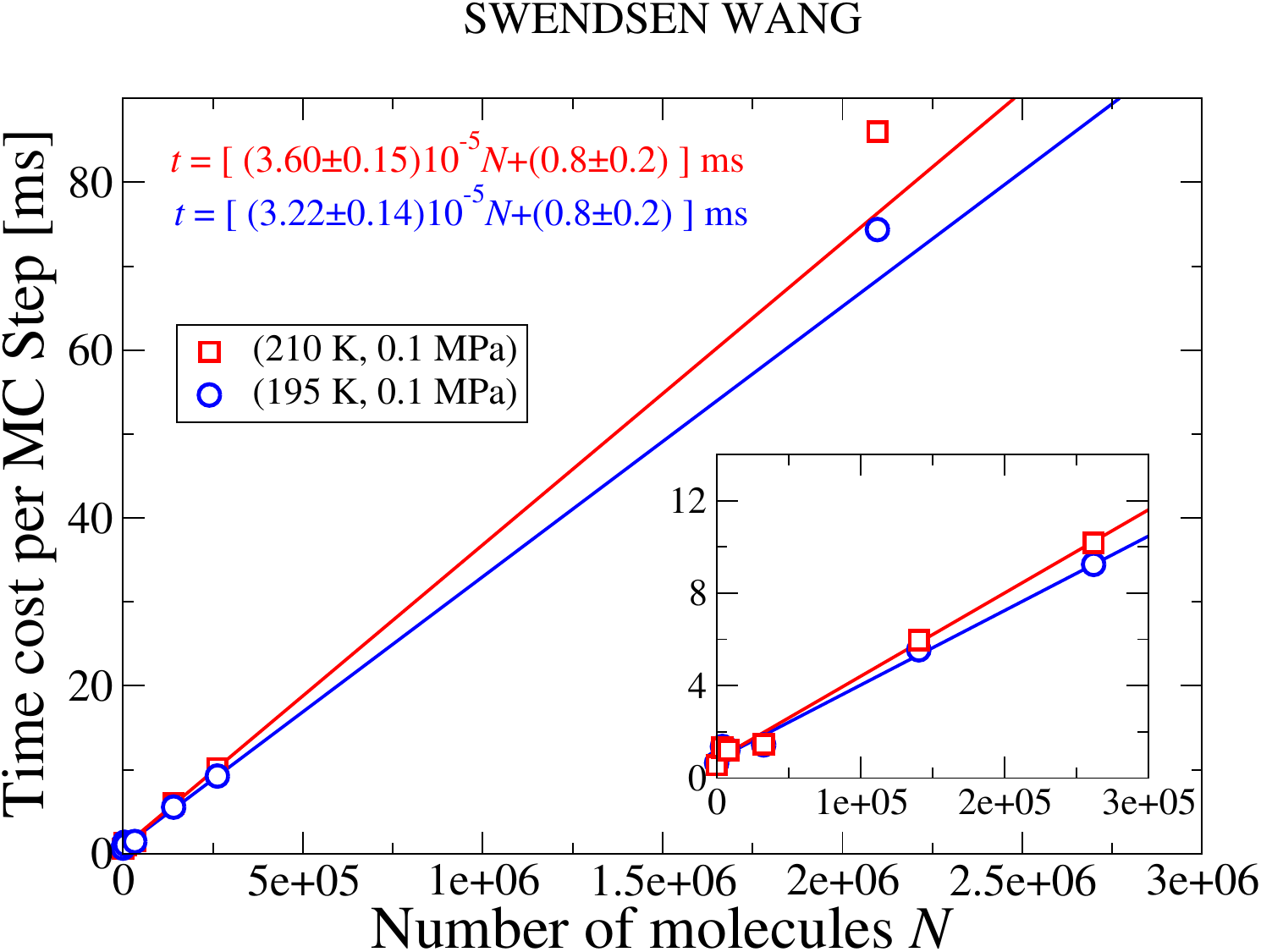}
\caption{{\bf Time cost of a parallel SW cluster update of $\mathbf{N}$ water molecules}. 
Red squares correspond to $T=$ 210 K (non-percolating cluster) and blue circles correspond to $ T = 195$ K (percolating cluster) at pressure $P=0.1$ MPa. 
Lines with matching colors indicate linear fits of the data, expressed as $ t = aN + b$, within the range 
$16 \leq N \leq 
2\,097\,152$.  
The fitting parameters
Are $a = (3.60 \pm 0.15) \times 10^{-5}$~ms and $ b = (0.8 \pm 0.2)$~ms for the red line, and
$a = (3.22 \pm 0.14) \times 10^{-5}$~ms and $ b = (0.8 \pm 0.2)$~ms for the blue line, respectively.
We observe that the cost increases faster than linear at large $N$ and exceeds that of the Metropolis MC, limiting our ability to explore sizes with tens of millions of water molecules, which contrasts with the Metropolis case.
Inset: The enlarged view at small $N$ highlights excellent linearity for the small $N$.
In both the main panel and the inset, error bars are smaller than the size of the symbols.
}
 \label{fig:SwendsenWangGPUtime}
\end{figure}

We find that the time cost of the parallel SW algorithm increases linearly for $N\leq 262,144$, although data at small values of $N$ are noisy (Fig.~\ref{fig:SwendsenWangGPUtime}). As with the parallel Metropolis algorithm, we attribute this to the suboptimal usage of GPU resources. 

At larger values of $N$, we observe additional time costs compared to linearity. As in the Metropolis case, we attribute this to limited resources for storing large arrays, such as those used for random numbers. However, the cost value reached at $N=2\,097\,152$ for SW is approximately 1.6 times larger than in the Metropolis case for the same size, which limits our ability to explore sizes with tens of millions of water molecules.

We note that the parallel SW update is faster under percolating conditions than in the absence of percolation. Although a better performance for a cluster algorithm is expected when the correlation length is large, because larger clusters lead to fewer in number, this result is not obvious. One might expect that the total time cost of the update is governed by the time cost of labeling the largest cluster, as seen in the sequential implementation. 

A possible explanation is that the analysis function converges rapidly irrespective of the cluster size, making the size of the largest cluster less relevant. Therefore, the difference in time cost between percolation and non-percolation likely stems from less efficient memory readings of the label array in smaller clusters by the scanning and labeling functions. 

A further consequence of this feature of the parallel implementation on GPUs is that the speedup factor relative to the sequential implementation on CPUs is greater under percolation conditions (Table \ref{table:SwendsenWangSpeedupFactor} and Fig.~S4). In particular, we find that for $N\geq 32\,768$, the SF under percolating conditions is nearly twice that under non-percolating conditions.

\begin{table*}
\centering
\caption{As in Table \ref{table:MetropolisSpeedupFactor}, but for the GPU Swendsen-Wang algorithm under the two thermodynamic conditions shown in Fig.~\ref{fig:SwendsenWangGPUtime}.}
\begin{tabular}{ |c|c|c|}
\hline
\multicolumn{3}{|c|}{SW Speedup Factor (SF)} \\
\hline
Number of molecules $N$ & ($T=$195 K,$P=$0.1 MPa) & ($T=$210 K, $P=$0.1 MPa) \\
    & Percolating & Non-Percolating \\
\hline
 64 & 0.0617(12) & 0.052(12) \\
\hline
 4\,096 & 3.17(4) & 2.04(3) \\
\hline
8\,000 & 7.38(10) & 4.57(3) \\
\hline
 32\,768 & 30.3(5) & 16.57(13) \\
\hline 
 140\,608 & 41.5(3) & 20.7(2) \\
\hline
 262\,144 & 47.8(7) & 21.89(11) \\
\hline
 2\,097\,152 & 65.0(2) & 24.99(6) \\
\hline
\end{tabular}
\label{table:SwendsenWangSpeedupFactor}
\end{table*}

Based on our results, we observe that the time cost of a single MC update for a CVF water sample of $N=32\,768$ molecules (as discussed in sections \ref{dynamics_metropolis} and \ref{dynamics_cluster}) is $0.15$ ms using Metropolis. In the case of SW, the time cost is $1.9$ ms for percolating clusters, i.e., approaching the LLCP, and $2.0$ ms for non-percolating clusters, away from the critical region. 

Therefore, the SW algorithm is approximately ten times more costly than Metropolis for $N=32\,768$. This result suggests that the SW algorithm should be employed whenever the autocorrelation time $\tau$ obtained with SW is at least ten times shorter than the $\tau$ obtained with Metropolis. As we have discussed above, this occurs when the system approaches the Widom line (the maximum of the correlation length $\xi$) or the region of maxima of specific heat (the maximum of enthalpy fluctuations).

\section{Conclusion}
\label{section:conclusion}

In this work, we implement efficient parallel MC algorithms for the CVF model of bulk water. In particular, we design a Metropolis algorithm based on a layered partition scheme and adapt the label equivalence algorithm from Hawick \citep{Hawick2010b} and Kalentev \citep{Kalentev2011} for simulations using the SW algorithm. Our results show that when the correlation length of the HB network is small, the parallel Metropolis algorithm is more efficient than the SW. This efficiency arises because the Metropolis algorithm takes less time per update to perform memory and computation tasks. Specifically, we demonstrate that a single Metropolis update is roughly ten times faster than an update with the SW algorithm for $N=32\,768$.

However, the Metropolis dynamics suffer from slowing down when the correlation length $\xi$ of the HB network is large. This occurs when the thermodynamic conditions are close to the Widom line, for example, at ambient pressure and supercooled conditions, $T \lesssim 205$ K, or near the LLCP (174 $\pm$ 14 MPa, 186 $\pm$ 4 K) \citep{coronas2024phase} where $\xi$ eventually diverges. Thanks to simultaneous updates of correlated clusters, the SW algorithm avoids the critical slowing down of the dynamics, enabling efficient sampling under those conditions. Therefore, we conclude that the SW algorithm should be preferred when the system is near a critical point or the corresponding Widom line, as the increased computational time for a single update is balanced by the fewer Monte Carlo steps required to yield statistically independent configurations. 

Furthermore, we observe that the speedup factor of the GPU implementation, in relation to the CPU implementation, of the two MC algorithms can be approximately 137 for Metropolis and 65 for SW when $N=2\, 097\, 152$. Therefore, regardless of the algorithm used, we find that GPU parallelization enables the CVF model to scale for simulations of unprecedentedly large water systems, reaching tens of millions of water molecules. 

For instance, we benchmark systems of 17\,576\,000 water molecules using the Metropolis algorithm, and 2\,097\,152 molecules for the SW cluster MC. The smaller size for the SW algorithm results from its higher computational cost in terms of time and memory compared to Metropolis dynamics.

Combining these results with the observation that the CVF model is reliable, given its quantitative accuracy around ambient conditions \citep{Coronas-2024}, and is transferable at extreme thermodynamic conditions \citep{coronas2024phase}, we conclude that the CVF model is suitable for addressing problems in nanotechnology and nanobiology due to its accuracy, efficiency, and scalability. 
Furthermore, we observe that, although many relevant issues in these scientific areas occur at near-ambient conditions, the model's transferability at extreme conditions is essential for a better understanding of phenomena such as protein denaturation upon heating, cooling, pressurization, or depressurization \citep{BiancoPRL2015}. 

To further support these conclusions, we have demonstrated in preliminary work \citep{CoronasThesis} that CVF water enables us to calculate the free energy landscape of extensive biological systems that were previously simulatable only with implicit solvents. 
In particular, we examined the sequestration of superoxide dismutase 1 (SOD1) proteins into crowded bovine serum albumin (BSA) globular protein and Fused in Sarcoma (FUS) disordered protein environments \citep{Samanta2021}, as well as the shear-induced unfolding of the von Willebrand factor \citep{Languin-Cattoen2021}. Both cases were previously analyzed using the OPEP protein model with implicit solvent \citep{Timr2023}. 

In conclusion, the CVF represents a REST—reliable, efficient, scalable, and transferable—model for water and hydrated systems. Its innovative approach holds the promise to transform free energy calculations for large-scale nano-bio systems, paving the way for groundbreaking discoveries in the field.

\section*{Conflict of Interest Statement}

The authors declare that the research was conducted in the absence of any commercial or financial relationships that could be construed as a potential conflict of interest.

\section*{Author Contributions}

L.E.C.: Methodology, Software, Validation, Formal Anlaysis, Investigation, Data curation, Writing - Original Draft, Writing - Review \& Editing, Visualization. O.V.: Methodology, Software, Formal Anlaysis, Investigation, Writing - Original Draft, Writing - Review \& Editing. G.F.: Conceptualization, Methodology, Validation, Formal Anlaysis, Investigation, Resources, Data curation, Writing - Original Draft, Writing - Review \& Editing, Supervision, Project administration, Funding acquisition.

\section*{Funding}
This work was supported by the Spanish Ministerio de Ciencia e Innovaci\'on/Agencia Estatal de Investigaci\'on [grant number MCIN/AEI/ 10.13039/ 501100011033]; the European Commission “ERDF A way of making Europe” [grant number PID2021-124297NB-C31]; the Universitat de Barcelona [grant number 5757200 APIF\_18\_19].

G.F. acknowledges the support from the Ministry of Universities 2023-2024 Mobility Subprogram within the Talent and its Employability Promotion State Program (PEICTI 2021-2023) and the Visitor Program of the Max Planck Institute for The Physics of Complex Systems for supporting a visit started in November 2022.

\section*{Acknowledgments}
We thank Valentino Bianco and Arne W. Zantop for their contributions to earlier versions of the CVF model.

\clearpage

\appendix

\section{Checkerboard partition for $\eta$ variables}
\label{CheckerboardPartition}

\begin{figure*}
	\centering
	(a) \includegraphics[scale=0.75]{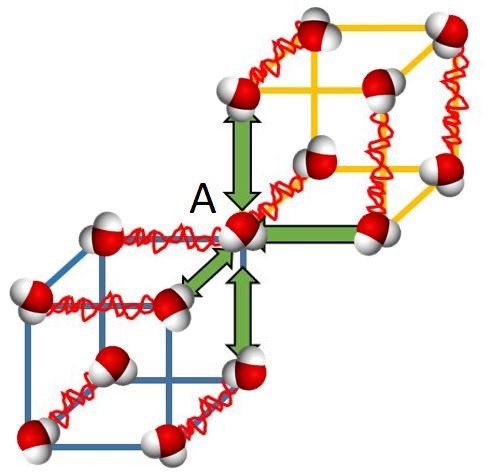}
	(b) \includegraphics[scale=0.65]{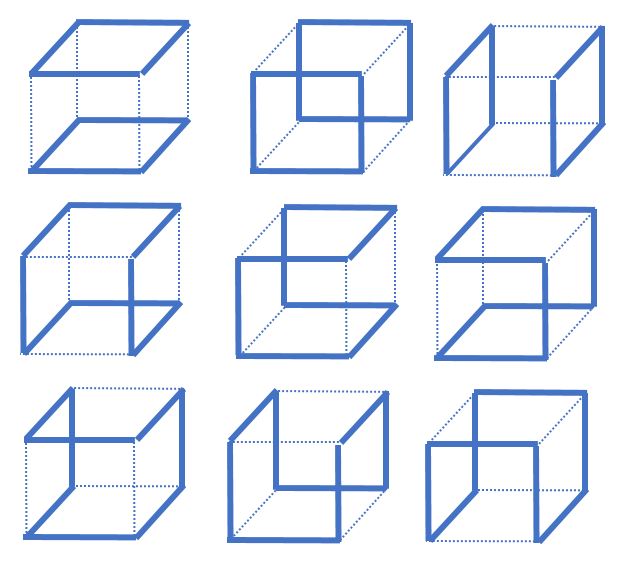}
	\caption{{\bf Example of checkerboard partition.} 
	{\bf (a)} The water molecule labeled "A" belongs to the blue and yellow checkerboard cubes (c.c.). Red broken lines represent forbidden ($\eta_{ij}=0$) edges, while all other (green, yellow, blue) edges correspond to $\eta_{ij}=1$. The molecule "A" can form a HB with the n.n. molecule $j$ when $\sigma_{Aj}=\sigma_{jA}$ along the edges marked by green double arrows. 
The sub-graph	$\mathcal{G}_0$ is made of all the red edges, while the sub-graph  $\mathcal{G}_1$ of all the rest. 
	{\bf (b)} The nine possible allowed configurations of the c.c. that allow us to define Hamiltonian cycles over the entire lattice, with thick lines representing $\eta_{ij}=1$, and dotted lines $\eta_{ij}=0$ edges. The sub-graph $\mathcal{G}_1$ in panel (a) is made of the center and bottom-center configurations of edges in panel (b).}
\label{fig:cube_def}
\end{figure*}

We partition the 3D space into cells belonging to a cubic lattice, each containing a water molecule. 
The reciprocal (cubic) lattice is a graph $
\mathcal{G}(\mathcal{V},\mathcal{L})$, where each molecule $i$ is a vertex, $i \in \mathcal{V}$, and each vertex $i$ is linked to its nearest neighbor (n.n.) vertex $j$ by an edge, to which we associate a variable $\eta_{ij}$, with $\eta_{ij} \in \mathcal{L}$, $
\mathcal{L}$ being the set of edges (Fig.~\ref{fig:cube_def}a). 

Next, we split $\mathcal{L}$ into two disjoint sets: $\mathcal{L}_0$ consisting of all the edges $\{\eta_{ij} = 0\}$ and $\mathcal{L}_1$ consisting of the remaining edges $\{\eta_{ij} = 1\}$. We use these sets to define the graphs $\mathcal{G}_0(\mathcal{V},\mathcal{L}_0)$ and $\mathcal{G}_1(\mathcal{V},\mathcal{L}_1)$. The goal is to generate sets $\mathcal{L}_1$ so that all vertices (water molecules) $i$ are connected to exactly four n.n. 

We achieve this by letting $\mathcal{G}_1$ be made of Hamiltonian cycles~\citep{Diestel2017}. A Hamiltonian cycle is a path that visits all the vertices once. In other words, if we start from any vertex $i$, we can follow the edges in $\mathcal{L}_1$ to visit all the vertices once and eventually return to the initial vertex. We note that if $\mathcal{G}_1$ is a Hamiltonian cycle, then all the molecules have exactly two variables $\eta_{ij}=0$ and four $\eta_{ij}=1$, which is the desired result (Fig.~\ref{fig:cube_def}b).

Next, we develop an algorithm that generates valid sets $(\mathcal{L}_0,\mathcal{L}_1)$ at every MC step (update of the $\eta_{ij}$ variables). The algorithm is specifically designed to suit the CVF model and can be easily parallelized. 

We adopt the checkerboard partition of $\mathcal{G}$, such that each molecule belongs to two, and only two, cubes (Fig.~\ref {fig:cube_def}a). Then, we set the edges of each checkerboard cubes (c.c.) to any of the nine predefined configurations shown in Fig.~\ref {fig:cube_def}b. For our previous considerations, the resulting  $\mathcal{G}_1$ is a Hamiltonian cycle. 

Because the state of a c.c. can change independently of any other c.c., this algorithm is intrinsically parallel. For its implementation, we assign one c.c. to a single GPU thread. We note that this algorithm requires a lattice with periodic boundary conditions and a linear size that is a multiple of four in each direction.

\section{Parallel code workflow}
\label{workflow}

In CUDA applications, tasks are distributed between the GPU (device) and the CPU (host), depending on whether they are parallelized or not. Arrays that are read/written by both must be allocated in both locations. To distinguish them, the usual criterion is to add the ${\tt dev\_}$ prefix if it is allocated on the device. To improve application performance, communication between the host and device should occur only when necessary.

\begin{figure*}
 \centering
 \includegraphics[width=0.8\textwidth]{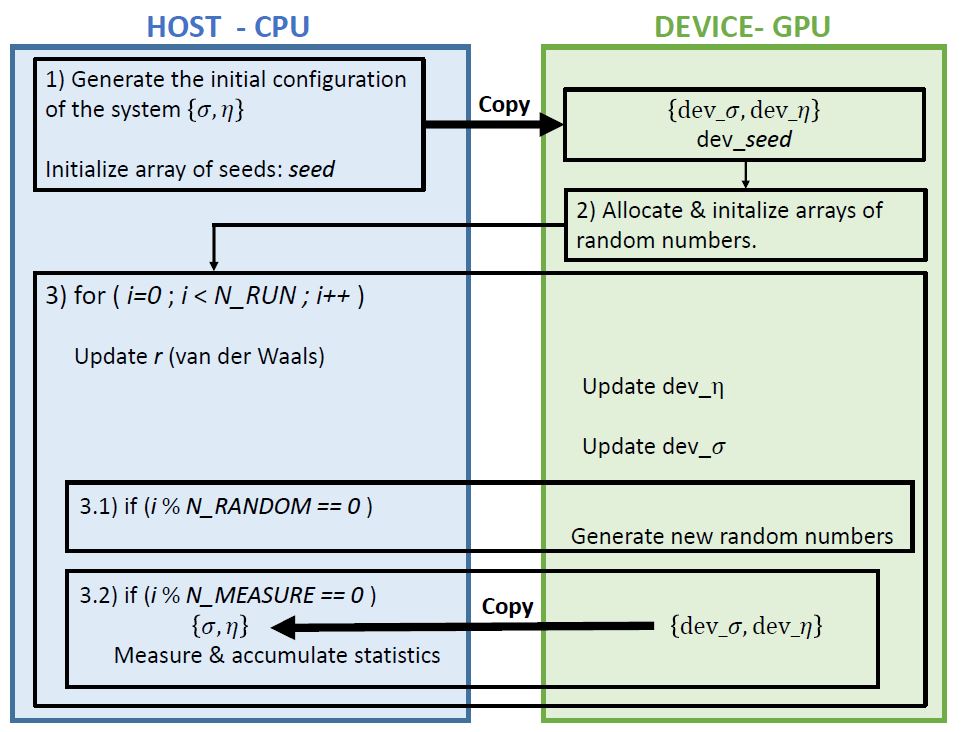} 
 \caption{Schematic workflow of the parallel CVF model program.}
 \label{fig:CPU-GPU_workflow}
\end{figure*}

We present the workflow scheme of our application in Fig. ~\ref{fig:CPU-GPU_workflow}. We begin generating the initial configuration of the system $\{\sigma,\eta\}$ on the host by initializing a vector of seeds that the CUDA random number generator will use. Next, we allocate the vectors ${\tt dev\_}\sigma$, ${\tt dev\_}\eta$, and {\tt dev\_seed} on the device and copy the values stored on the host. Then, we generate arrays of random number vectors on the device using the $ {\tt cuRand} $ library. These arrays contain the random numbers that will be read by the CUDA kernels, following the principle of deferred decisions. In the Appendix~\ref{random_numbers}, we provide more details on random number generation and their usage. The main loop performs ${\tt N\_RUN}$ MC steps (updating $r$, $\eta$, and $\sigma$) as described in section \ref{MonteCarloStep}. 

\section{Generation and usage of random numbers}
\label{random_numbers}

Random numbers are required for various tasks, including proposing new $\sigma_{ij}$ and $\eta_{ij}$ states, or deciding whether to accept or reject a new configuration. In sequential MC, one typically generates a random number only when necessary. However, this approach is inconvenient in parallel algorithms, as it is preferable to avoid thread divergence.

In our parallelization scheme, we adopt the \textit{principle of deferred decisions} strategy. For each task that requires a random number, we allocate an array of random numbers of size $N\cdot{\tt N\_RANDOM}$. This array contains the maximum number of random numbers required for ${\tt N\_RANDOM}$ executions of the kernels, where each kernel launches a thread per molecule. We index the arrays of random numbers according to ${\tt rnd\_idx} = r\cdot N+ {\tt cell}$, where $0\leq r < {\tt N\_RANDOM}$. Then, for a given calculation that involves a random number, each thread reads its corresponding random number, taking advantage of memory coalescing. To avoid thread divergence, the computation of random numbers is performed even if they are not needed later.

We generate random numbers on the GPU using the {\tt cuRAND} library. In particular, we produce arrays of $N$ uncorrelated sequences of random numbers and overwrite them every {\tt N\_RANDOM} steps, once they have been used.

\section{Workstation}
\label{workstation}

We measure the performance of our parallel Metropolis and SW algorithms (on combined GPU and CPU) and compare them to the results obtained from a sequential implementation on the CPU. The results were obtained using a workstation with an AMD Ryzen 7 5800H processor running at 4.4 GHz and an NVIDIA GeForce RTX 3060 with 6 GB of global memory and 3840 CUDA cores. The CUDA Toolkit version used is 11.5.1 (released in November 2021).

\section{Speedup factors}
\label{speedup}

For an accurate comparison between sequential and parallel algorithms, the measurement of the time cost of the GPU algorithm includes the time cost of ${\rm GPU}\to{\rm CPU}$ memory transactions (Appendix~\ref{workflow}), as well as the GPU time spent on generating random numbers (Appendix~\ref{random_numbers}). The time cost of the CPU algorithm includes only the CPU time spent updating the system, as there are no memory transactions. We define the speedup factor (SF) as the ratio between the time cost in CPU and GPU, $SF \equiv t^{\rm CPU}/t^{\rm GPU}$. Thus, the GPU-parallel algorithm is more efficient than the CPU-sequential one if ${\rm SF}>1$.





\begin{figure*}
\includegraphics[width=\textwidth]{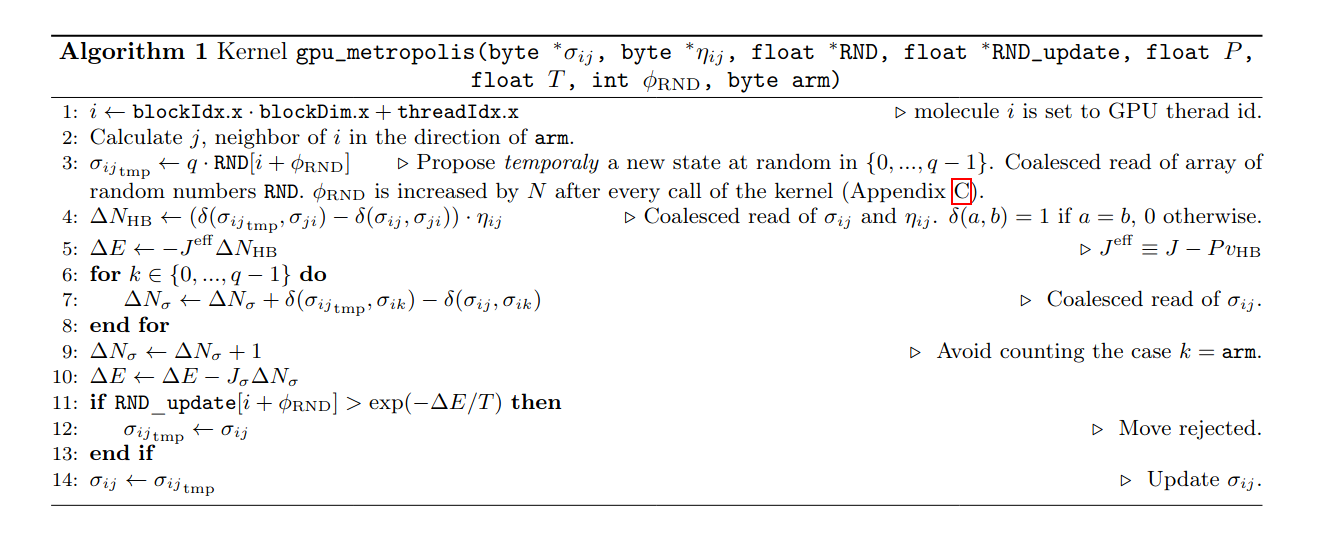}
\end{figure*}

\begin{figure*}
\includegraphics[width=\textwidth]{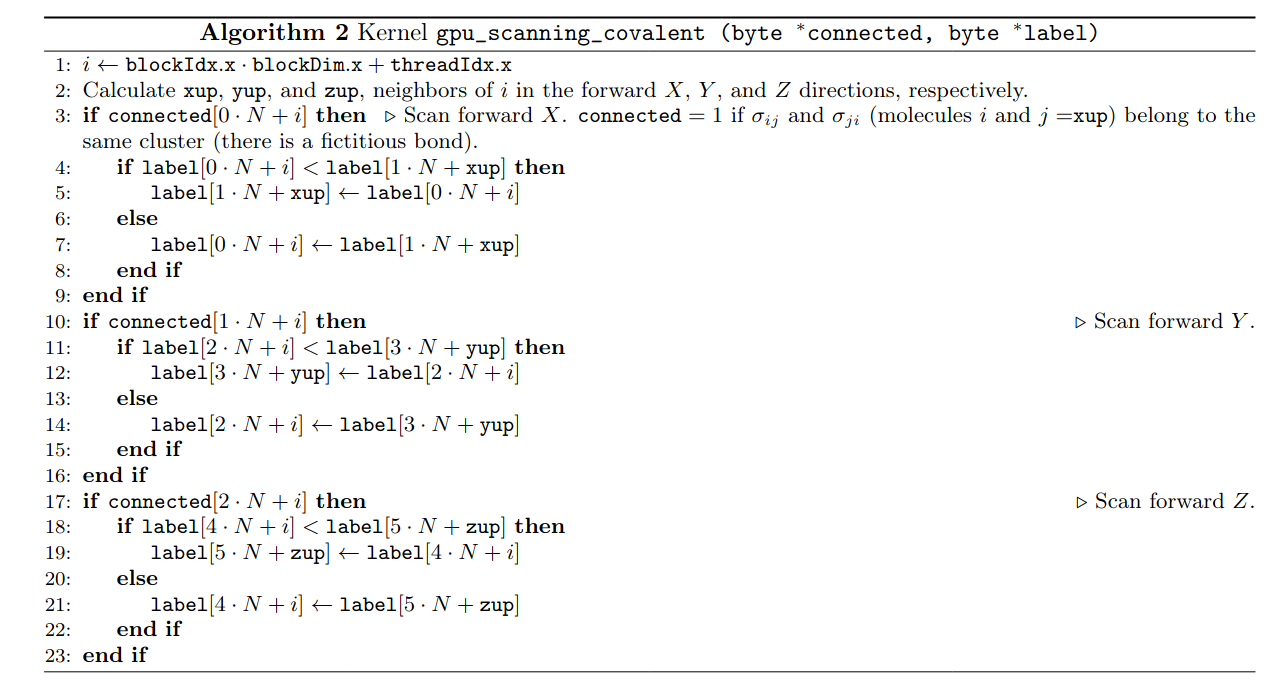}
\end{figure*}

\begin{figure*}
\includegraphics[width=\textwidth]{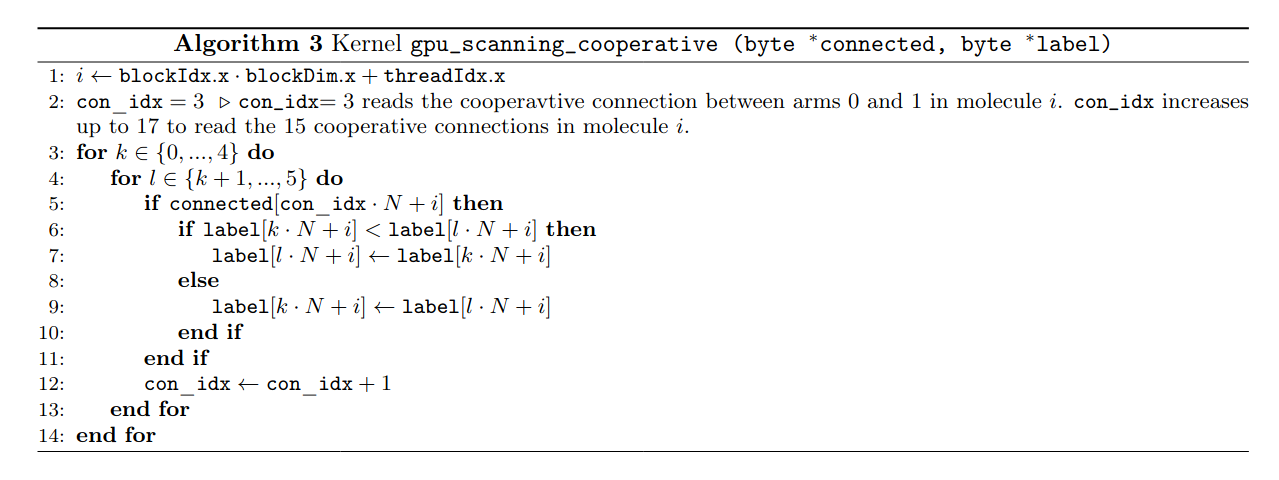}
\end{figure*}

\begin{figure*}
\includegraphics[width=\textwidth]{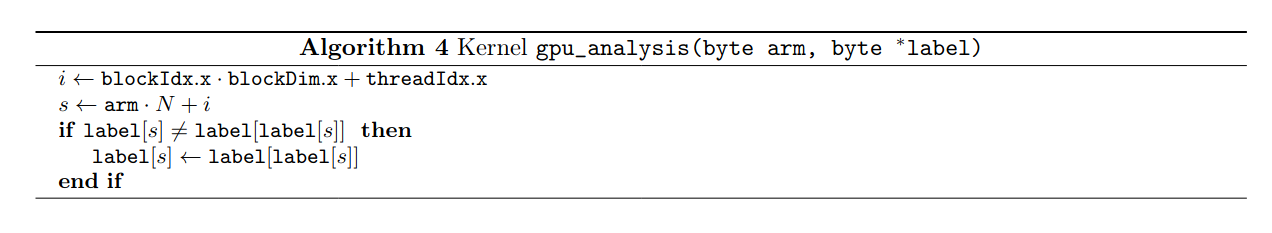}
\end{figure*}

\clearpage

\begin{figure*}
    \centering
    \includegraphics[scale=0.5]{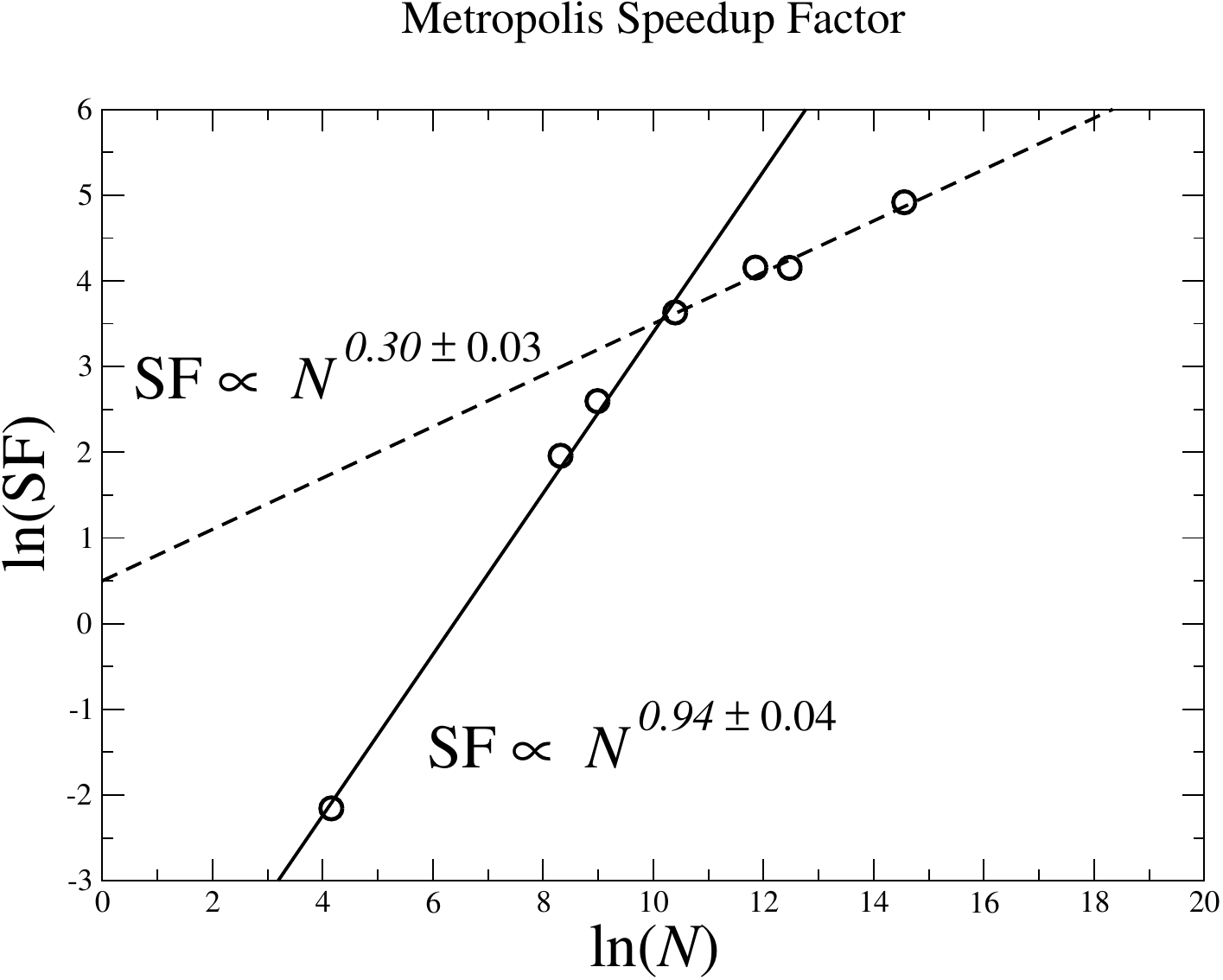}
    \caption{{\bf The scaling of the Metropolis speedup factor $\mathbf{SF(N) \propto N^\gamma}$ in the range $\mathbf{64 \leq N \leq 2\, 097\, 152}$ shows a crossover at $\mathbf{N=32\,768}$.} For small $N \leq 32\,768$, the exponent $\gamma = 0.94\pm0.04$ is close to $1$, indicating that the SF increases approximately linearly with $N$. We interpret this behavior as resulting from the fact that in this regime $t^{\rm GPU}$ is nearly constant, whereas $t^{\rm CPU}$ increases linearly. For $32\,768 \leq N \leq 2\,097\,152 $, the SF follows a power law with exponent $\gamma = 0.30\pm 0.03$. In this regime, $t^{\rm GPU}$ increases linearly (Fig. \ref{fig:MetroplisGPUtime}), while $t^{\rm CPU}$ deviates from linearity as the large $\sigma$ and $\eta$ arrays exceed the RAM storage capacity. Error bars are smaller than the size of the symbols.}
    \label{fig:MetropolisSpeedupFactor}
\end{figure*}

\begin{figure*}
    \centering
    \includegraphics[scale=0.5]{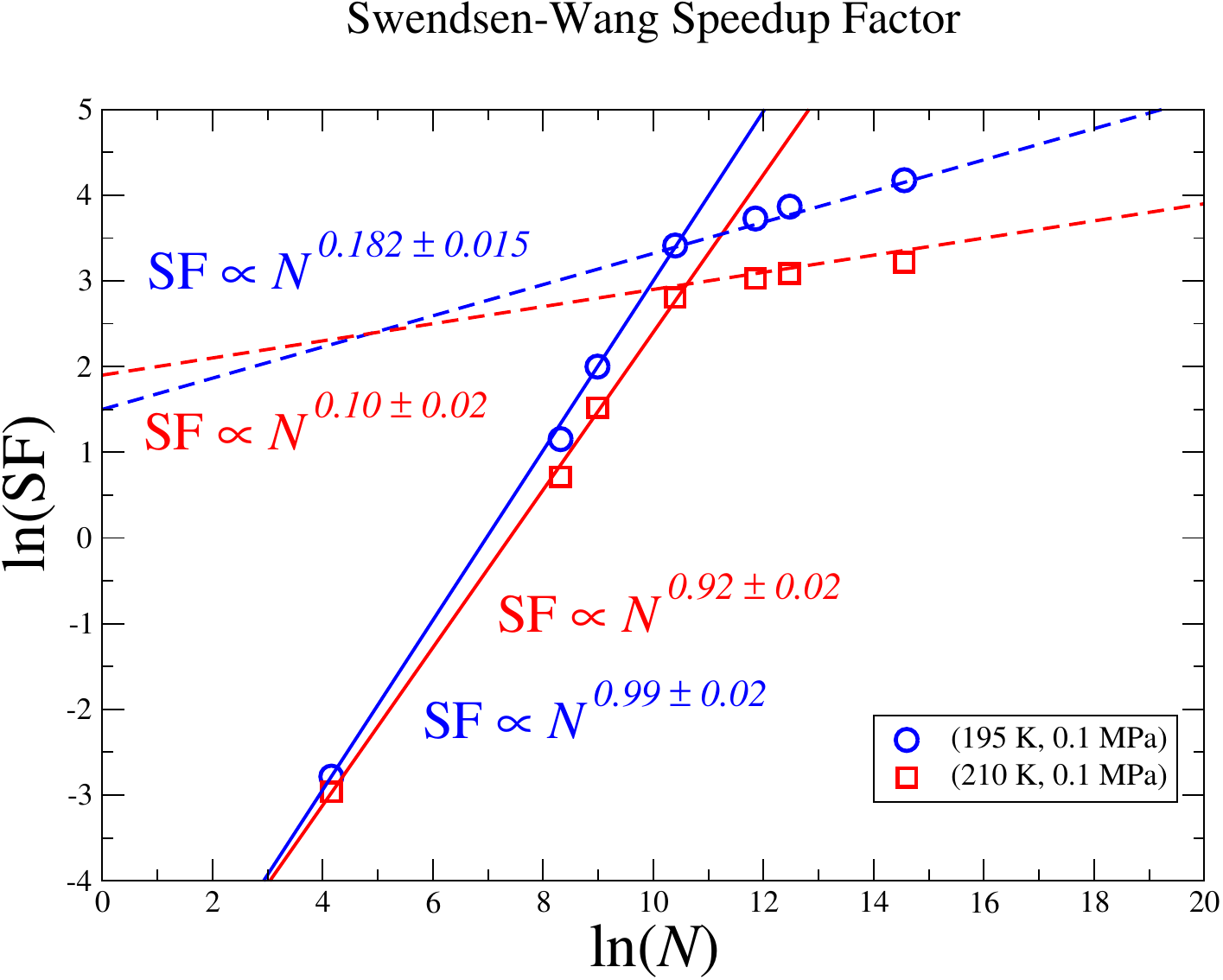}
    \caption{{\bf The scaling of the SW speedup factor $\mathbf{SF(N)\propto N^\gamma}$ in the range $\mathbf{64 \leq N \leq 2\, 097\, 152}$}. For both percolating (blue circles) and non-percolating (red squares) cases, the SF shows a crossover at $N=32\,768$ as in Metropolis. Under non-percolating conditions, the SF displays a crossover from $\gamma = 0.92\pm0.02$ to $\gamma = 0.10\pm0.02$. Under percolation, the crossover is from $\gamma = 0.99\pm0.02$ to $\gamma = 0.182\pm0.015$. Error bars are smaller than the size of the symbols.}
    \label{fig:SwendsenWangSpeedupFactor}
\end{figure*}

\end{document}